\documentclass[prb,twocolumn,superscriptaddress]{revtex4}
\usepackage{amsmath}
\usepackage{epsfig}
\usepackage{dcolumn}
\usepackage{bm}
\usepackage{slashed}
\usepackage[toc,page]{appendix}
\usepackage{float}
\usepackage{amsmath}
\usepackage{amssymb}
\usepackage{ulem}   
\usepackage{graphicx}

\def\gapp{\lower.35em\hbox{$\stackrel{\textstyle>}{\sim}$}}
\def\lapp{\lower.35em\hbox{$\stackrel{\textstyle<}{\sim}$}}

\newcommand{\bs}[1]{{\boldsymbol{#1}}}

\begin{document}
\bibliographystyle{apsrev}
%
\title{Interaction driven phases in the honeycomb lattice from exact diagonalization}

\author{Noel A. Garc\'ia-Mart\'inez}
\author{Adolfo G. Grushin}
\affiliation{Instituto de Ciencia de Materiales de Madrid, CSIC, Cantoblanco,
E-28049 Madrid, Spain}
\author{Titus Neupert}
\affiliation{Condensed Matter Theory Group, 
Paul Scherrer Institute, CH-5232 Villigen PSI,
Switzerland            } 
\author{Bel\'en Valenzuela }
\affiliation{Instituto de Ciencia de Materiales de Madrid, CSIC, Cantoblanco,
E-28049 Madrid, Spain}
\author{Eduardo V. Castro}
\affiliation{CFIF, Instituto Superior 
T\'ecnico, TU Lisbon, Av. Rovisco Pais, 1049-001 Lisboa, Portugal}

\begin{abstract}
We investigate the fate of interaction driven phases in the half-filled honeycomb lattice for finite systems via exact diagonalization with nearest and next nearest neighbour interactions.
We find evidence for a charge density wave phase, a Kekul\'e bond order and a sublattice charge modulated phase in agreement with previously reported mean-field phase diagrams. No clear sign of an interaction driven Chern insulator phase (Haldane phase) is found despite being predicted by the same mean-field analysis. We
characterize these phases by their ground state degeneracy and by calculating 
charge order and bond order correlation functions. 
\end{abstract}

\maketitle

\section{Introduction}

The role of electron-electron interactions in graphene \cite{CNetal09r} has been a fruitful subject of research even before this material was discovered. \cite{GGV94} Although important progress has been made towards a full understanding of their effect, \cite{KUP12} there are still fundamental questions that need to be clarified. One of such open questions regards the fate of band electrons in graphene's honeycomb lattice when subject to repulsive interactions at half-filling. The plethora of techniques \cite{RQH08,CGV11,HML12,DMM12,HCW13} available to study the effect of interactions in this system has produced a range of interesting predictions. In particular, in a series of works, \cite{RQH08,WF10,CGV11,GCC13} several groups have produced compatible mean-field phase diagrams that suggest that electrons in a honeycomb lattice with extended Hubbard interactions at half filling stabilize a Chern insulator (CI) phase with topological character and quantized Hall conductivity. This phase is nothing but the celebrated Haldane phase \cite{H88} and it is realized at moderate value of the nearest neighbor (NN) $V_{1}$ and next to nearest neighbor (NNN) $V_{2}$ interactions but always with $V_{2}>V_{1}$. 

Interestingly, the CI phase is embedded in a rich structure of other competing orders in the phase diagram. The first of these is a charge density wave order (CDW) \cite{RQH08} at $V_{1}\gg V_{2}$ with charge imbalance between the two different sublattices that reduces the amount of NN interaction energy to be paid. A sublattice charge modulated phase (CMs) \cite{GCC13} was also found for $V_{2}\gg V_{1}$ with charge imbalance over the same sublattice that compensates for the large $V_{2}$ cost.  At sufficiently large $V_{2}\sim V_{1}$, a Kekul\'e bond order emerged \cite{C00,HCM07,WF10} characterized by a $Z_3$ order parameter which can lead to fractionalized excitations of $\pm e/2$ at the long wavelength limit. \cite{HCM07}  These phases have the additional interest of being also examples of an interaction driven gap for low energy quasiparticles in the honeycomb lattice. Together, these works provide a clear consistent picture of the possible phases available within the mean-field perspective. However, 
the results are subject to the limitations of mean field theory, for (i) there is no certainty that all local order parameters relevant to the low energy physics have been considered,~\cite{GCC13} (ii) the ground state of the system might not be adiabatically connected to mean-field state with a local order parameter and (iii) the mean field phase can be over/under estimated in the parameter region of the phase diagram.\\
To test the mean-field picture it is necessary to employ different tools as independent checks for the presence of the mean-field phases.  One of such tools is exact diagonalization (ED) which we explore in this work. It is based on the ED of the Hamiltonian for finite lattice sizes and it provides, in principle, an unbiased analysis of interactions. 
The main limitations for ED in two-dimensional quantum systems are the smallness of system sizes that can be studied. Finite size effects might well out-range the energy scale of a potential many-body gap of incompressible ground states, so that the incompressibility cannot be recognized.
Therefore, the limitations of ED and the mean-field approach are to a large extend complementary. If both methods yield the same phase for a region in parameter space, this provides strong evidence that the true ground state in the thermodynamic limit will be of this nature. Phases that cannot be easily detected with neither ED nor the mean-field approach include those with incommensurate long-range order. For example, depending on the system size and the particular geometry, ED might favor commensurate phases against frustrated phases and one has to be careful to explore (whenever possible) different sizes, and/or aspect ratios \cite{VSR10,VSR11} to pin down the relevant competing phases. Indeed, ED has proven useful in studies of the Haldane-Hubbard model\cite{VSR10,VSR11,WSZ10} and the $\pi$-flux model, \cite{JHZ13} complementing other techniques such as quantum Monte-Carlo and variational cluster approximation used in studies of the Hubbard and Kane-Mele-Hubbard model in the Honeycomb lattice. \cite{RL10,MLW10,HLA11,ZZW11,YXL11,HML12,WEL12,SOY12,HWGF13,HCW13,C13}\\

Motivated by these results, and in particular by the interaction driven phases found in existing mean-field calculations, in this work we study the spinless extended Hubbard model with both NN and NNN interactions in the honeycomb lattice at half filling via ED of small finite size systems. We will investigate and characterize the phase diagram for electronic phases that are driven by Coulomb interactions in the honeycomb lattice as an independent check for the mean-field picture. We will provide evidence for the appearance of some of the phases that were previously obtained in mean-field calculations. These include the CDW, the Kekul\'e bond order and the CMs phases which surround a trivial semi-metal (SM) phase. Surprisingly, for the studied lattice sizes, we find no clear sign of the previously reported interaction driven CI phase. As for the phases that do appear, we will characterize them by their ground-state degeneracy and by computing the charge density and bond order correlation functions. 

In section \ref{sec:model} we introduce the model and establish notation conventions. 
In section \ref{sec:PD} we present the complete phase diagram of the honeycomb lattice at half filling with NN and NNN interactions. 
We will discuss the main properties and characterize each of the appearing phases. In section \ref{sec:hald} we relate our findings with previous works and discuss the absence the interaction driven CI phase. Finally, in section \ref{sec:conc} we summarize our main findings.

\section{\label{sec:model} The model}
\begin{figure}
\includegraphics[scale=0.25]{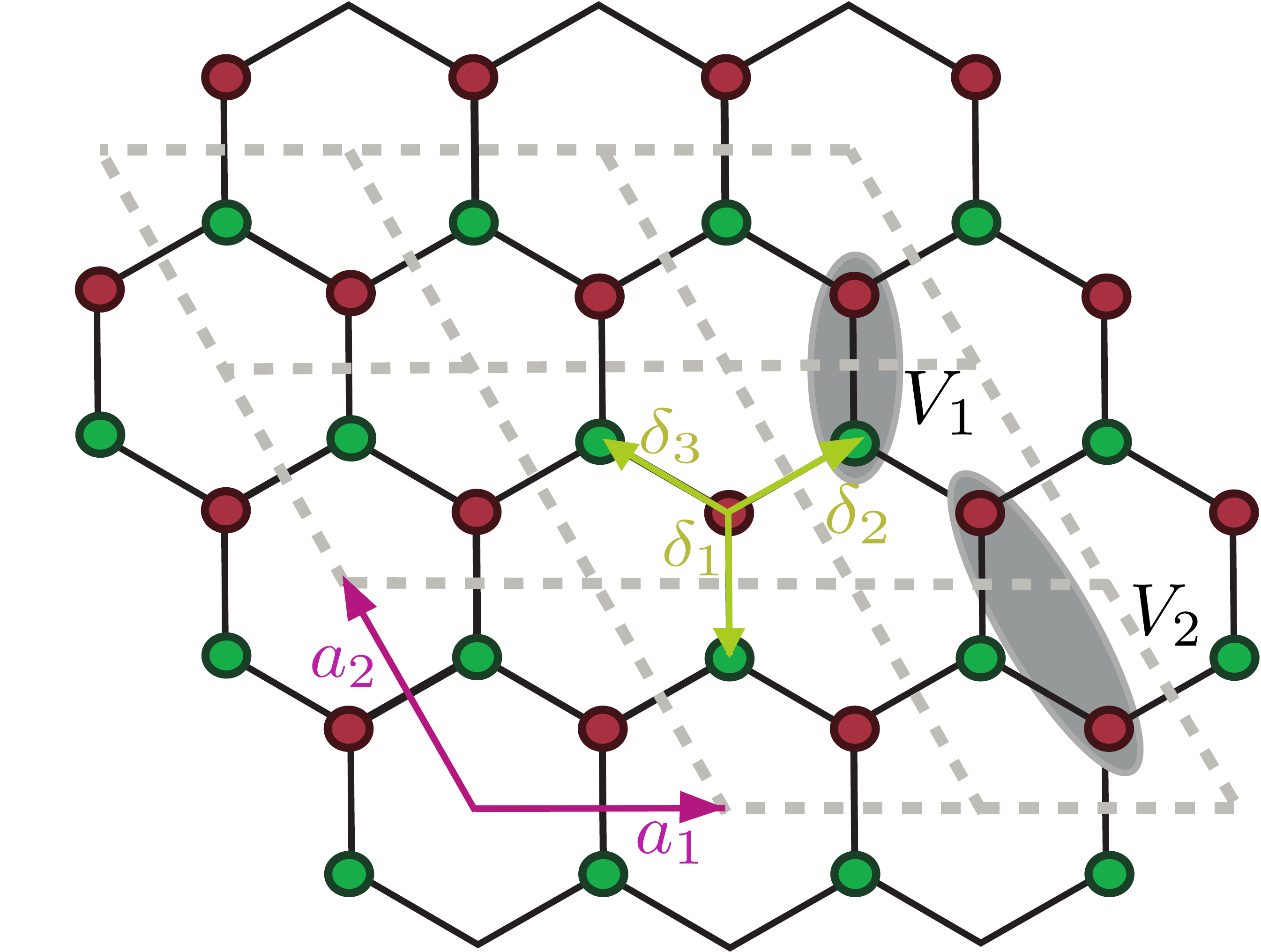}
\caption{\label{Fig:Defs} (Color online)
The extended Hubbard model on the honeycomb lattice. 
A and B sublattices are represented by green and red circles. 
The basis vectors and nearest neighbours vectors defined in the text
are $\bs{a}_{1,2}$ and $\bs{\delta}_{1,2,3}$ respectively. 
The NN and NNN interactions $V_{1}$ and $V_{2}$
are represented by grey ellipses. The grey dotted lines enclose 
the $\Omega=3\times3$ cluster with periodic boundary conditions.}
\end{figure} 
We start with the spinless extended Hubbard model for electrons in a honeycomb lattice with nearest neighbor (NN) interaction $V_{1}$ and next to nearest neighbor (NNN) interaction $V_{2}$. The Hamiltonian in real space reads:
\begin{eqnarray}
H:=-t\sum_{\left\langle i,j\right\rangle }(c^{\dagger}_{i}c_{j}+h.c.)
\;+
V_{1}\sum_{\left\langle i,j\right\rangle }n_{i}n_{j}+
V_{2}\sum_{\left\langle \left\langle i,j\right\rangle \right\rangle }n_{i}n_{j}\,
\label{eq:H}
\end{eqnarray}
where $t$ is the nearest neighbor hopping and $c_{i}$  
annihilates an electron at the $i$-th site of the honeycomb lattice. 
Each of the two triangular sublattices A and B is spanned by the basis vectors
$\bs{a}_{1}=\bs{\delta}_{2}-\bs{\delta}_{3}$ and 
$\bs{a}_{2}=\bs{\delta}_{3}-\bs{\delta}_{1}$ defined through the three nearest neighbors $\bs{\delta}_{1}=a(0,-1)$,  
$\bs{\delta}_{3}=a(\sqrt{3}/2,1/2)$ and $\bs{\delta}_{3}=a(-\sqrt{3}/2,+1/2)$ as shown in Fig.~\ref{Fig:Defs}.
Transforming to Fourier space by defining $a_{\bs{k}}^{\dagger}:=\frac{1}{\sqrt{\Omega}}\sum_{i\in A}c_{i}^{\dagger}e^{i\bs{k}.\cdot\bs{r}_{i}}$
and $b_{\bs{k}}^{\dagger}:=\frac{1}{\sqrt{\Omega}}\sum_{i\in B}c_{i}^{\dagger}e^{i\bs{k}.\cdot\bs{r}_{i}}$ the Hamiltonian \eqref{eq:H} can be expressed as
\begin{equation}
\begin{split}
H  =&-t\sum_{\bs{k}}\gamma_{\bs{k}}a_{\bs{k}}^{\dagger}b_{\bs{k}}+\mbox{h.c.} \\
 & +\frac{V_{1}}{\Omega}\sum_{\bs{k},\bs{k}',\bs{q}}\gamma_{\bs{q}}a_{\bs{k}}^{\dagger}a_{\bs{k}-\bs{q}}
 b_{\bs{k}'}^{\dagger}b_{\bs{k}'+\bs{q}}+\mbox{h.c.}+ \\
 & +\frac{V_{2}}{\Omega}\sum_{\bs{k},\bs{k}',\bs{q}}\chi_{\bs{q}}\left(a_{\bs{k}}^{\dagger}a_{\bs{k}-\bs{q}}
 a_{\bs{k}'}^{\dagger}a_{\bs{k}'+\bs{q}} + b_{\bs{k}}^{\dagger}
 b_{\bs{k}-\bs{q}}b_{\bs{k}'}^{\dagger}b_{\bs{k}'+\bs{q}}\right)\,,
 \label{eq:Hfourier}
 \end{split}
 \end{equation}
where $\gamma_{\bs{k}}=(1+e^{i\bs{k}\cdot\bs{a}_{2}}
+e^{i\bs{k}\cdot(\bs{a}_{1}+\bs{a}_{2})})$ and
$\chi_{\bs{k}}=(e^{i\bs{k}\cdot\bs{a}_{1}}
+e^{i\bs{k}\cdot\bs{a}_{2}}+e^{i\bs{k}\cdot(\bs{a}_{1}
+\bs{a}_{2})})$ are NN and NNN form factors respectively 
and $\Omega$ is the number of unit cells. \\

In what follows we investigate the phase diagram as a function of $V_{1}$ and $V_{2}$ via  ED of small clusters of size $\Omega=3\times3$ [see Fig.~\ref{Fig:Defs}] and $\Omega=3\times4$ using periodic boundary conditions. 

\section{\label{sec:PD}Phase diagram}

\begin{figure}
\begin{minipage}{.45\linewidth}
\begin{center}
\includegraphics[scale=0.30]{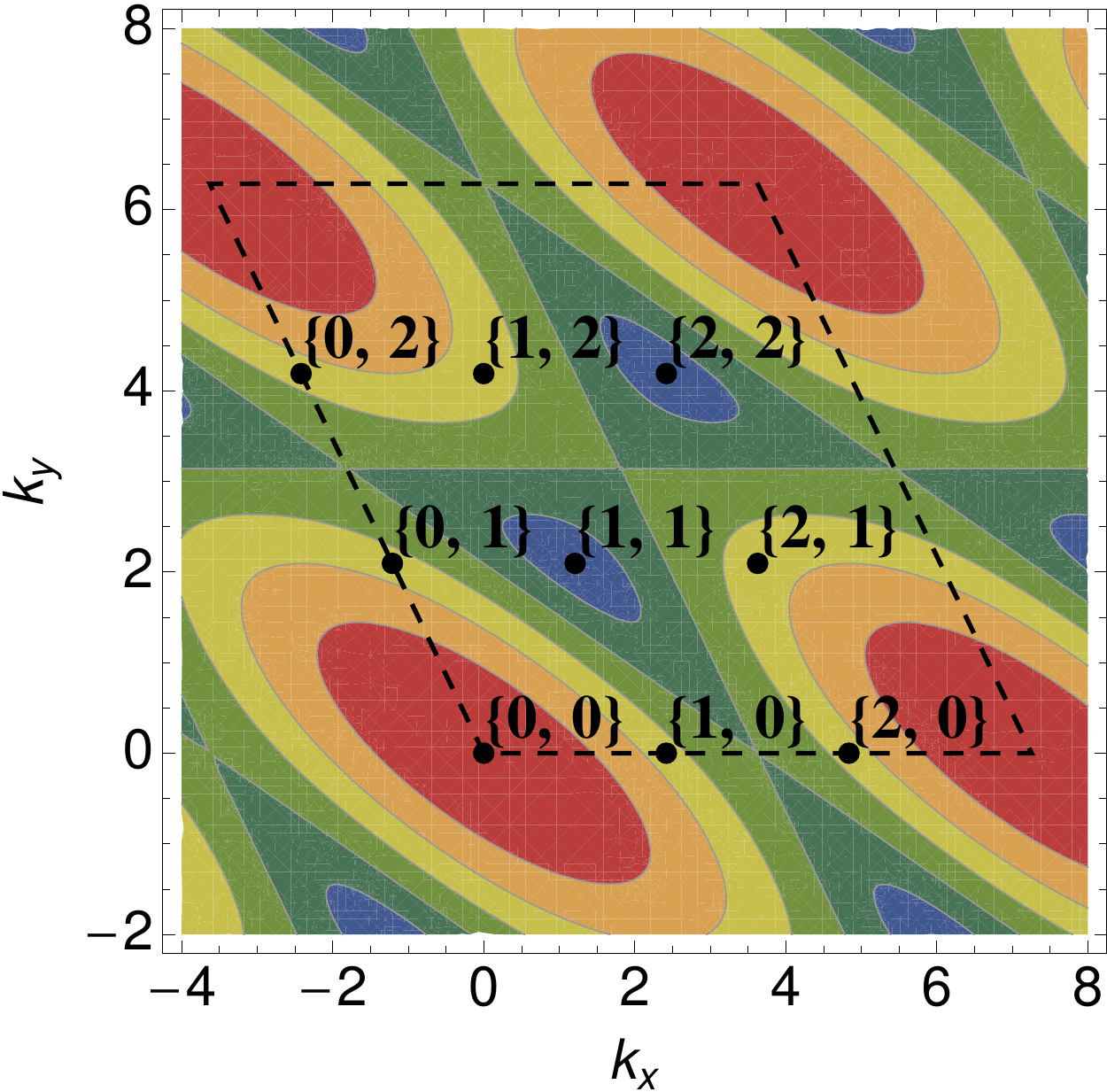}
\begin{center}
(a) 
\end{center}
\end{center}
\end{minipage}
\begin{minipage}{.45\linewidth}
\begin{center}
\includegraphics[scale=0.30]{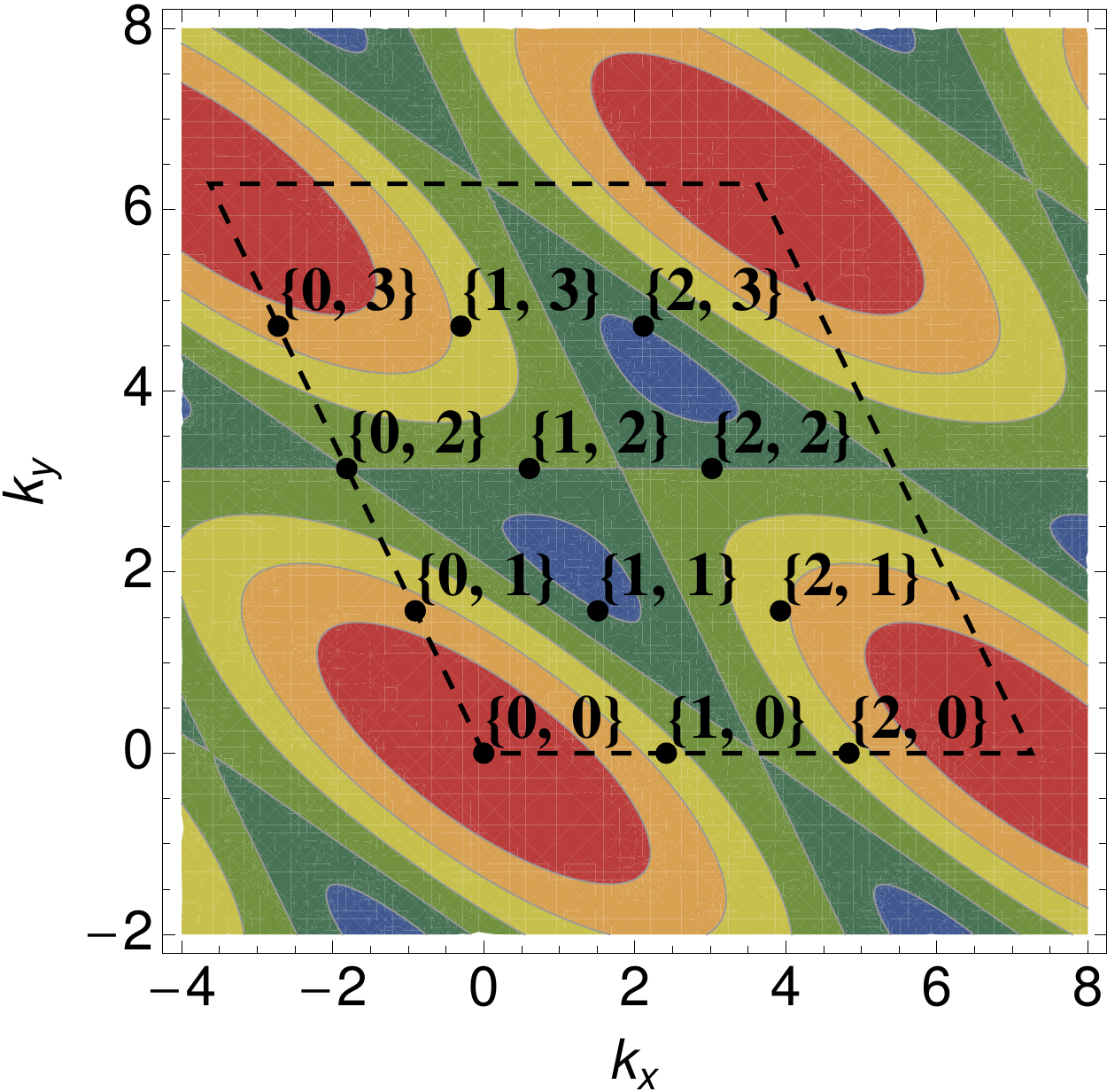}
\begin{center}
(b) 
\end{center}
\end{center}
\end{minipage}
\caption{\label{Fig:nonintbands} (Color online) Contour plot of the non-interacting band structure obtained from \eqref{eq:H} with $V_{1}=V_{2}=0$. Superimposed is the discretized BZ (dashed line) for a lattice with (a) $\Omega = 3\times 3$ and (b) $\Omega = 3\times4$ unit cells. The inset text shows the momentum label $\bs{Q}=(Q_1,Q_2)$ (see text).}
\end{figure}  

In order to implement the ED of Hamiltonian~\eqref{eq:Hfourier} 
we discretize the Brillouin zone (BZ) with a 
lattice $\Omega=L_1\times L_2$ of points that span the BZ area.
The band structure of graphene and two different set of lattice sizes
 are shown in Fig. \ref{Fig:nonintbands}.
For the $\Omega=3\times3$ lattice 
there are $9$ points per band in the BZ 
each to be filled with one electron.
There is one at the 
$\bs{\Gamma}$ point, two at the $\bs{K}$ and 
$\bs{K'}$ points, and a set of six energetically 
degenerate points. For a given particle $i$, we label its 
momentum by its coordinates in momentum space 
$k^{(i)}_1,k^{(i)}_2$ or, alternatively, with the discrete 
one-dimensional integer label $Q^{(i)}=k^{(i)}_{1}+L_1 k^{(i)}_{2}$. 
In this notation and for this lattice the 
$\bs{\Gamma}$ point corresponds to momentum $(0,0)$, or $Q^{(i)}=0$, and the $\bs{K}$ 
and $\bs{K'}$ points are at $(1,1)$, or $Q^{(i)}=4$, and $(2,2)$, or $Q^{(i)}=8$, 
respectively [see Fig. \ref{Fig:nonintbands}(a)]. 
In general $k^{(i)}_1\in \left[0,L_1-1\right]$, 
$k^{(i)}_2 \in \left[0,L_2-1\right]$ and $Q^{(i)} \in \left[0,L_1L_2-1\right]$.\\

\begin{figure*}
\begin{minipage}{.49\linewidth}
\begin{center}
\includegraphics[scale=0.40]{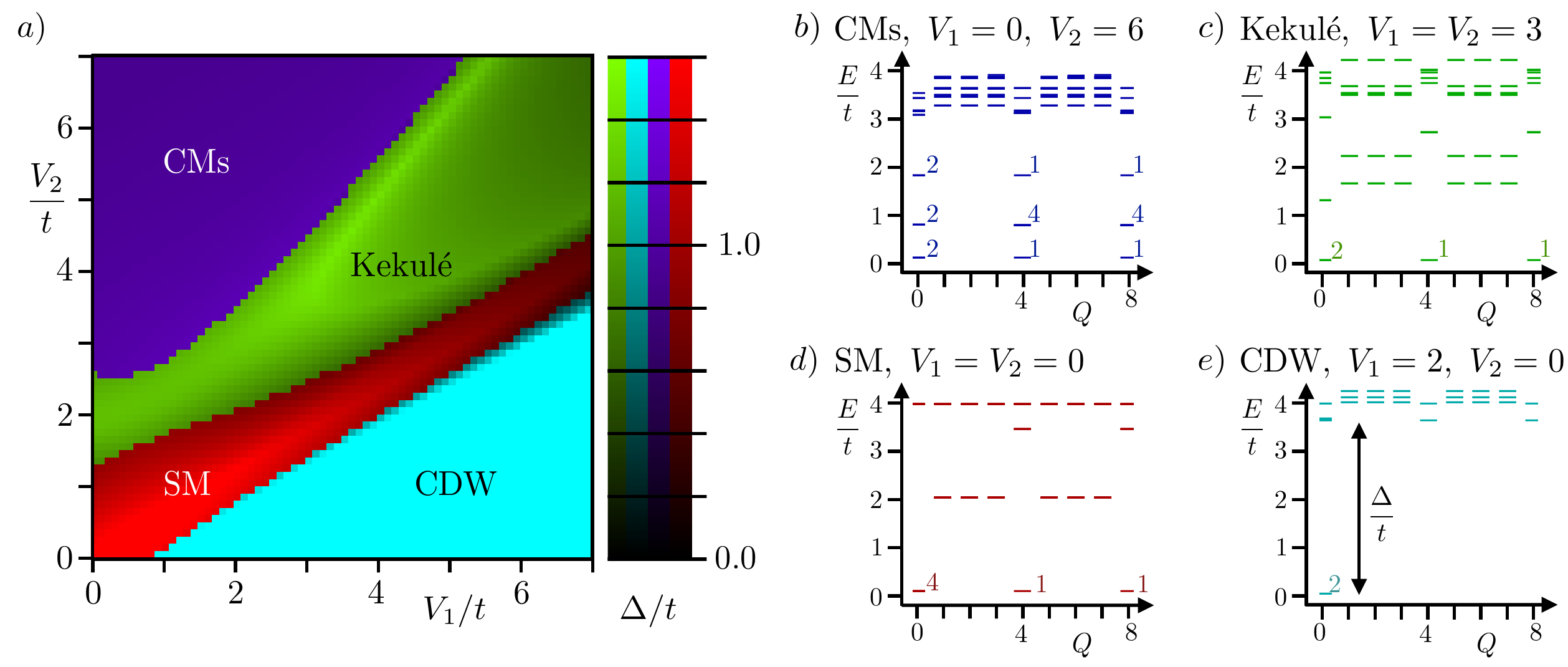}
\begin{center}
(i) 
\end{center}
\end{center}
\end{minipage}
\begin{minipage}{.49\linewidth}
\begin{center}
\includegraphics[scale=0.20]{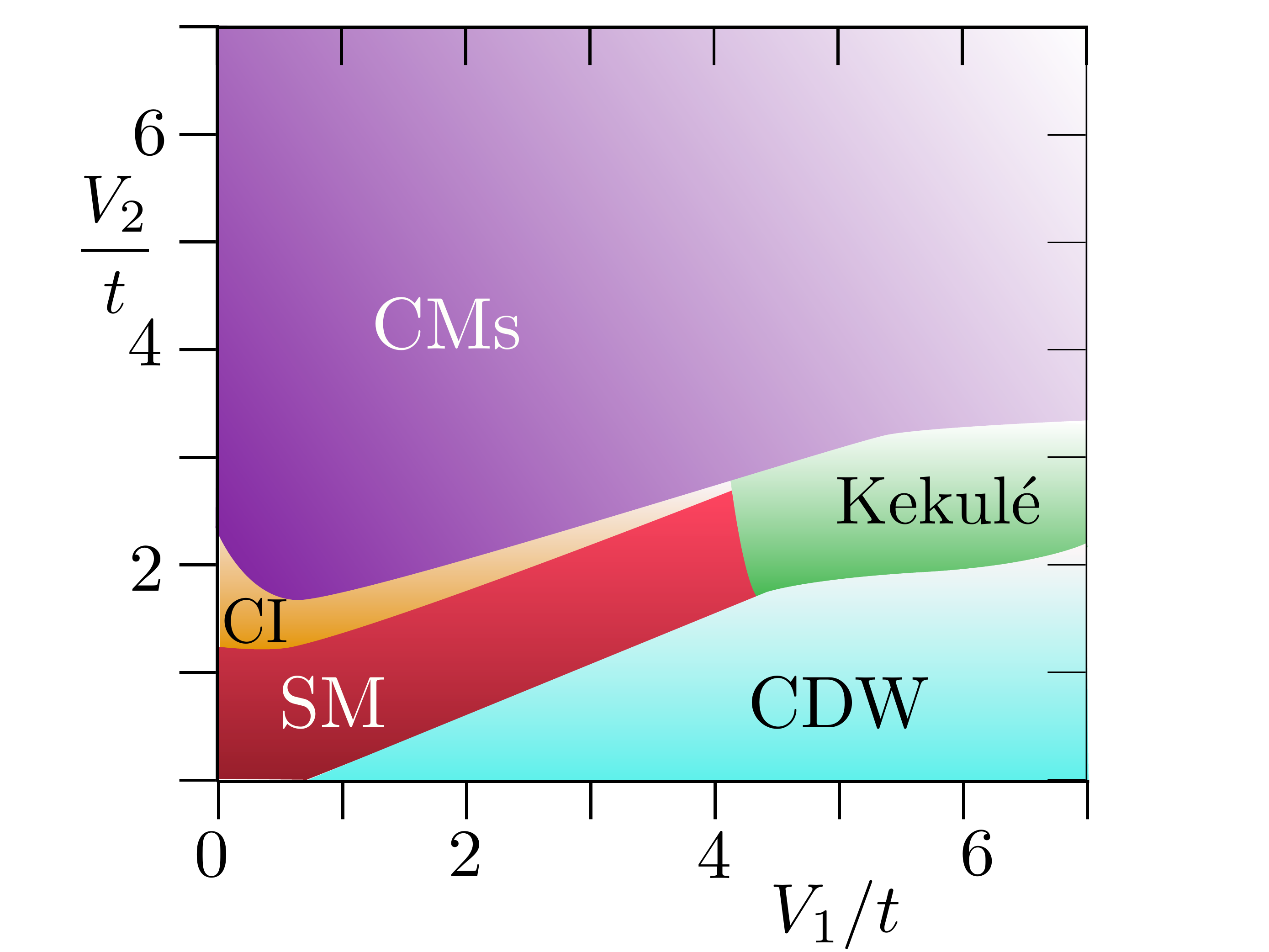}
\begin{center}
(ii) 
\end{center}
\end{center}
\end{minipage}
\caption{\label{Fig:PD3x3} (Color online) (i) ED Phase diagram at $n=1/2$ for a $\Omega=3\times 3$ system. 
The brightness of each color is proportional to the size of the many-body gap $\Delta/t$.
The right hand side shows the energy spectrum against total momentum
 $Q=Q_{1}+L_{1}Q_{2}$ for the (b) CMs phase, (c) Kekul\'e phase
 (d) SM phase and (e) CDW phase. The small numbers indicate the degeneracy of each state. The zero of energies is chosen to be the ground state energy. The phases are identified by the number of ground states over which there is the highest gap. (ii) Mean-field phase diagram calculated following Ref. \onlinecite{GCC13}.}
\end{figure*}

Since the interaction in Hamiltonian~\eqref{eq:Hfourier} conserves the total momentum,
in ED we can diagonalize 
independently each total momentum sector subspace $Q=\sum_{i} Q^{(i)}$ 
with $Q \in \left[0,L_1L_2-1\right]$, where the momentum is defined modulo $\Omega$. Therefore all eigenvalues and eigenvectors that we obtain are labelled by $Q$.
The phase diagram for $\Omega=3\times 3$ with $N=9$ 
particles (i.\ e., $\nu=N/(2\Omega)=1/2 $ filling) and 
representative eigenvalue spectra as a function of $Q$ 
are shown in Fig.~\ref{Fig:PD3x3}(i). By focusing on the ground 
state degeneracy we identify a phase by the number of ground states 
over which there is the highest gap. In what follows we will distinguish and characterize the
four distinct phases. We will argue that they correspond to the 
SM, Kekul\'e, CDW, and CMs phases and discuss their main signatures. We note that the phase boundaries might be altered 
by going to larger systems or applying alternative definitions to identify the phases.

In the following section we will use these findings to relate 
to previous works to finally compare with the mean-field diagram in Fig.~\ref{Fig:PD3x3}~(ii) of Ref. \onlinecite{GCC13} 
which includes all possible (non-superconducting) mean-field decouplings with a tripled unit cell. In particular, this phase diagram is consistent with past mean-field studies for which the absence of the Kekul\'e \cite{RQH08,DMM12} or CMs phases \cite{RQH08,WF10,CGV11,DMM12} in the mean field phase diagrams was due to the fact that these works did not allow
for these mean-field solutions.

\subsection{Semi-metal phase}

This phase, labelled SM and shown in red in Fig. \ref{Fig:PD3x3}(a) is straight forward to 
characterize since it stems from the noninteracting ($V_{1}=V_{2}=0$) limit of Hamiltonian~\eqref{eq:Hfourier}. 
For $\Omega=3\times3$ at half filling ($N=9$)
there are $2\Omega=18$ lattice sites to fit $9$ particles. Seven of them sit at the lower 
states, one at the $\bs{\Gamma}$ point at $(0,0)$ and
six particles go to the six degenerate momenta at $(1,0),(2,0),(0,1),(2,1),(0,2),(1,2)$.
We have two particles left for four degenerate single-particle states, two at the $\bs{K}$ point 
and two at the $\bs{K}'$ point, since at these points there
are two degenerate states, one from each band. 
This gives a freedom to choose the ground state. 
We have $2$ particles to fill $4$ states, 
the degeneracy of which is given by the binomial coefficient $C(4,2)=6$ 
which is the ground-state degeneracy for the non interacting case.
Out of these six possibilities, four of them have a particle at $\bs{K}$ 
and a particle at $\bs{K'}$ and thus a total momentum of $Q=0$. 
The remaining two configurations have two particles at the same valley. 
Having both at $\bs{K}=(1,1)$ results in a total momentum $(2,2)$ or $Q=8$. 
Similarly placing the two last particles at $\bs{K'}=(2,2)$
we expect to have a single state at momentum $(1,1)$, or $Q=4$.

To summarize, the non-interacting Hamiltonian in ED has a sixfold quasi degenerate 
ground-state at half-filling with four states at $Q=0$, one state at $Q=4$ 
and one state at $Q=8$. 

We observe this structure for a finite region of parameters 
colored red in Fig. \ref{Fig:PD3x3}(a) connected to the non-interacting Hamiltonian
and thus we interpret this phase as a SM phase. 
The spectrum for such phase is shown in Fig. \ref{Fig:PD3x3}(d) where the six-fold 
quasi degenerate ground-state are observed at the momenta discussed above.

\subsection{Charge Density Wave phase}

\begin{figure}[h]
\includegraphics[scale=0.15]{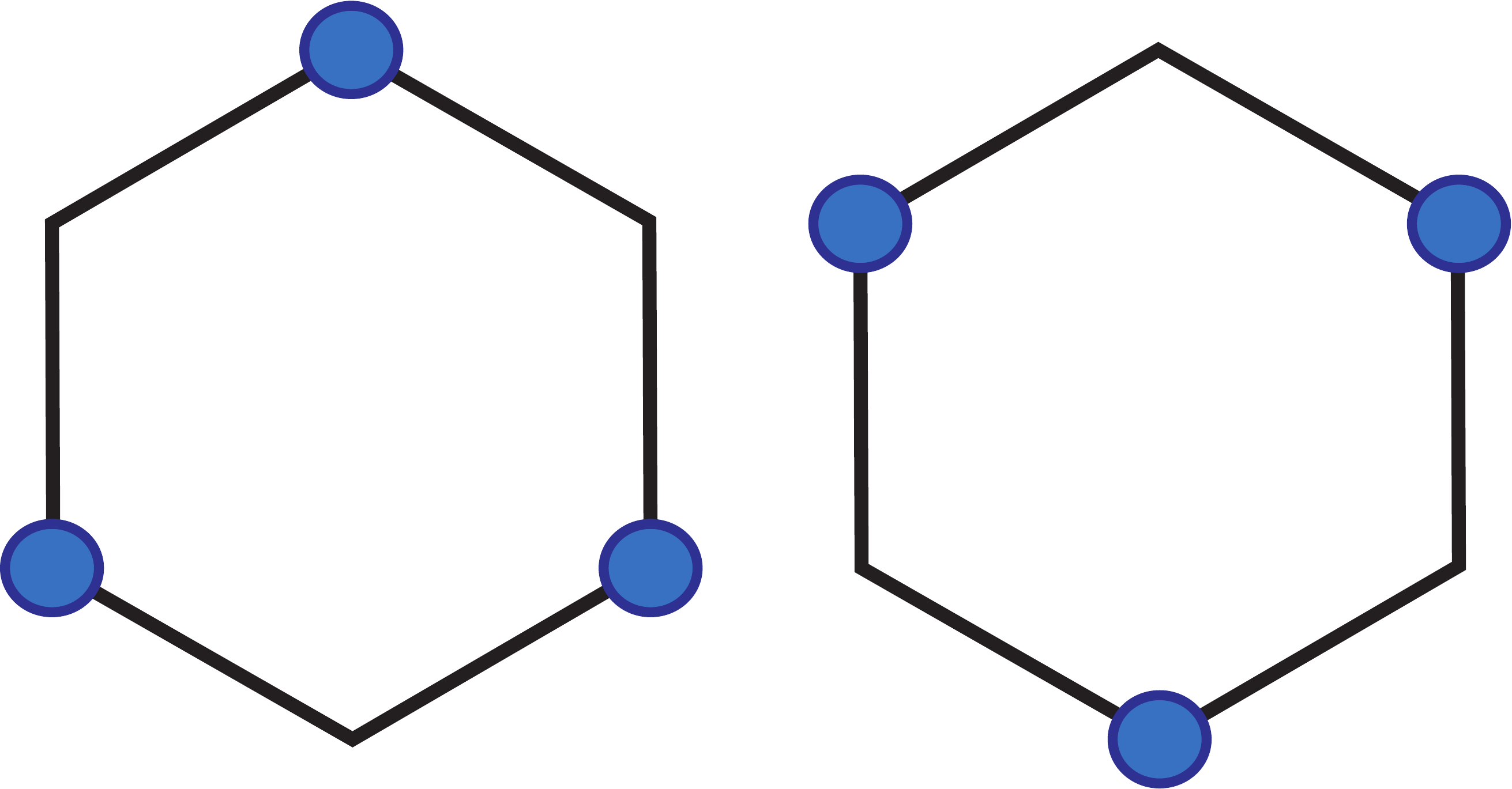}
\caption{\label{Fig:CDWpatt} (Color online) CDW pattern at $V_{1}/t\to\infty$ with $V_{2}=0$.}
\end{figure}

The second phase that we identify is labelled CDW and shown in light blue in Fig. \ref{Fig:PD3x3}(a). Its spectrum shows a two-fold degenerate ground state at $Q=0$ [Fig. \ref{Fig:PD3x3}(e)] and would break spontaneously the sublattice symmetry in the thermodynamic limit. The most transparent way to understand that this phase is indeed a CDW is to investigate the strong coupling limit at $V_{1}/t\to\infty$ with $V_{2}=0$ to which this phase is adiabatically connected. Calculating the degeneracy of such a strong coupling state is a classical problem, the ground state of which is represented in Fig.~\ref{Fig:CDWpatt}.

As only one sublattice is occupied in either of these classical ground states, both of them are zero-energy eigenstates of the NN interaction $V_{1}$.
We expect this state to appear at total momentum $Q=0$ since it is a charge density wave (CDW) order
state within the unit cell. Indeed, the ED of the Hamiltonian with $V_{1}\neq 0$ and $V_{2}=t=0$ yields exactly this two fold degenerate ground-state at zero energy. The excited states that can be computed classically also coincide both in energy and degeneracy in this limit.

It is possible to connect this strong coupling phase to the two-fold degenerate phase shown in Fig.~\ref{Fig:PD3x3} simply by increasing the hopping continuously to see that both phases are indeed connected without ever closing the many-body gap. From this fact alone we can already conclude that this state is a CDW state.
A further check of this picture comes from calculating the charge density wave 
modulation in the same spirit as described in Ref.~\onlinecite{GNC12}. Suppose that we have a phase with a set of (quasi-)degenerate ground states $\left | m \right >$, $m=1,\cdots, N_{\mathrm{gs}}$. For these, we define the sublattice-staggered electron density matrix 
\begin{equation}\label{densop}
\begin{split}
\rho^{mm'}_{\bs{r}} := &\dfrac{1}{\Omega}\sum_{\bs{k}}
 e^{i\left(\bs{Q}_{m}-\bs{Q}_{m'}\right)\cdot \bs{r}}
 \\
&\times
\left\langle m\right|
a^\dagger_{\bs{k}+\left(\bs{Q}_{m}-\bs{Q}_{m'}\right)}a_{\bs{k}}
-b^\dagger_{\bs{k}+\left(\bs{Q}_{m}-\bs{Q}_{m'}\right)}b_{\bs{k}}
\left|m' \right\rangle.
\end{split}
\end{equation} 
For the CDW case we have in particular that $m=1,2$  with both $\bs{Q}_{1}=\bs{Q}_{2}=0$. Note that since $\bs{Q}_{1}=\bs{Q}_{2}$ it is not possible to build a linear combination of ground states that generates charge modulation outside the unit cell which is consistent with the CDW we are trying to characterize. 

We now diagonalize the $2\times 2$ matrix $ \rho_{\bs{r}}$ for representative points inside the CDW phase. This generates two  $\bs{r}$-independent eigenvectors and eigenvalues, $\bs{v}_{i}$ and $\lambda_{i}$ ($i=1,2$). The former represent the two independent super-positions of the two-fold degenerate ground states while the latter represent the two possible sublattice imbalances. Therefore if finite, the eigenvalues are the defining feature of the CDW phase. For example, for $V_{1}=5t$ and $V_{2}=0$ we find that $\lambda_{1}=-0.99$ and $\lambda_2=0.97$. The former (latter) corresponds to a state with particles localized in sublattice B (A) as represented in Fig.~\ref{Fig:CDWpatt} confirming the CDW interpretation of the state.  \\
Finally, a transparent way to understand this state is to relate it with the non-interacting honeycomb lattice with a staggered chemical potential $\pm m$ in the $A$($B$) sublattice, i.e. the non-interacting version of the CDW state. Upon filling the band structure for this simple case at half-filling and for $\Omega=3\times 3$ the two particles highest in energy have momenta $\bs{K}$ and $\bs{K'}$, thus corresponding to a single state of total momentum $Q=0$. Within the interacting model and since symmetry breaking is absent for finite systems, by ED of the interacting Hamiltonian we find both $\pm m$ and $\mp m$ configurations which then give a degeneracy of two.\\
\subsection{Kekul\'e phase}
\begin{figure*}
\begin{minipage}{.49\linewidth}
\begin{center}
\includegraphics[scale=0.25]{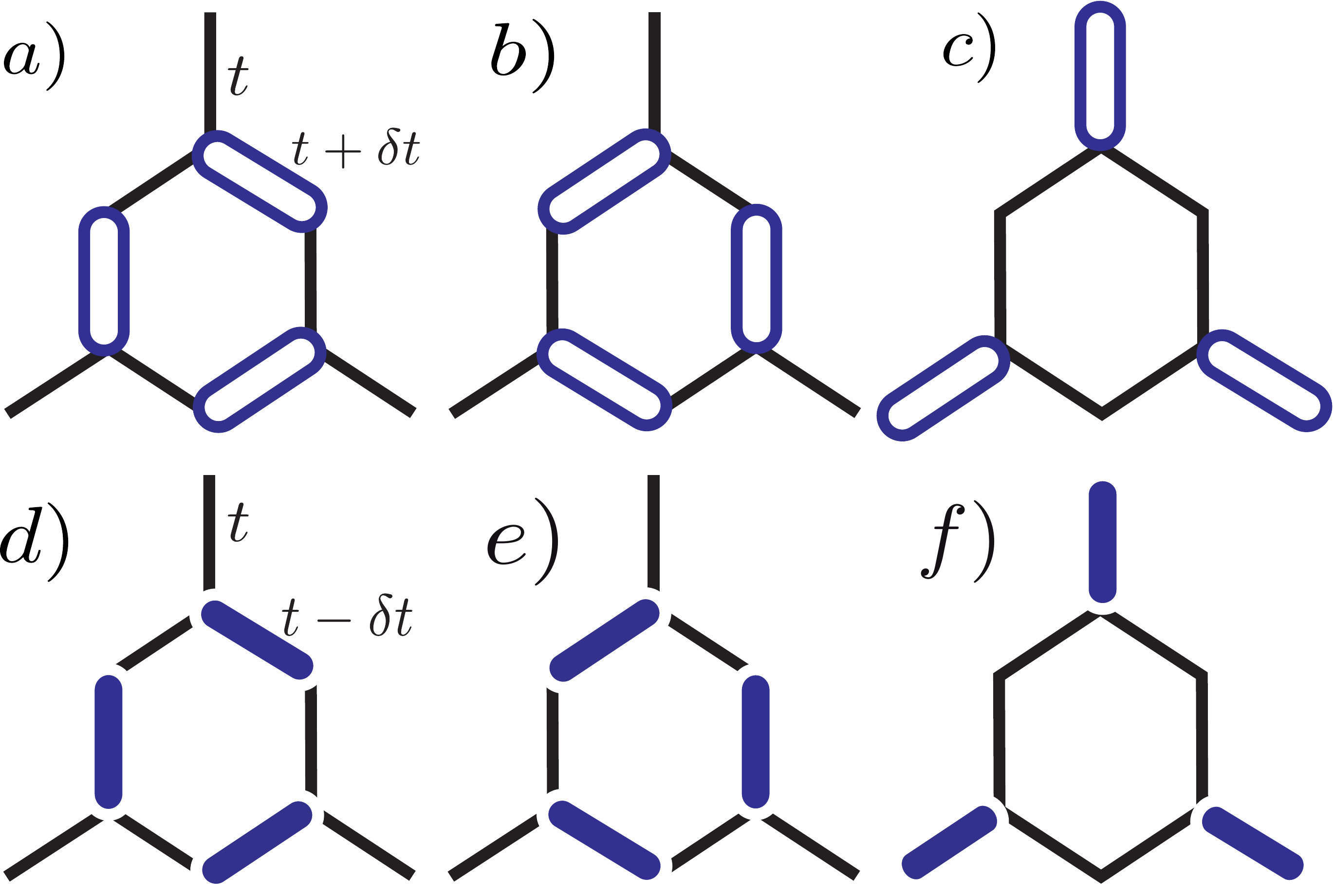}
\begin{center}
(i) 
\end{center}
\end{center}
\end{minipage}
\begin{minipage}{.49\linewidth}
\begin{center}
\includegraphics[scale=0.25]{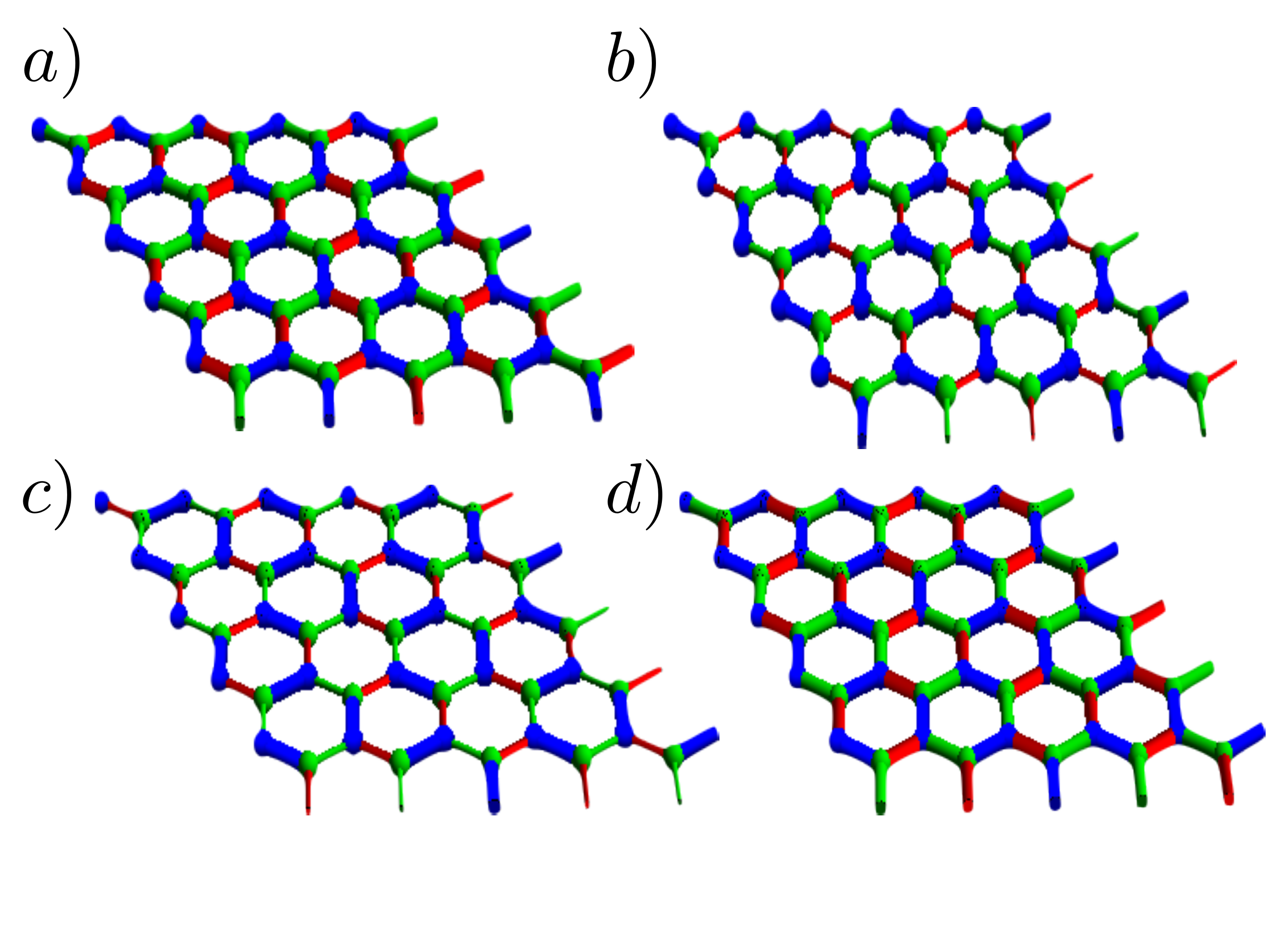}
\begin{center}
(ii) 
\end{center}
\end{center}
\end{minipage}
\caption{\label{Fig:Kekulepatt} (Color online) (i) The six different types of Kekul\'e orders. 
From these six patterns one can produce the
second row out of the first row only by
 including an overall homogeneous hopping decrease 
 d) + e) + f). Thus only four patterns are independent.
(ii) Four independent Kekul\'e super-positions obtained from \eqref{kekmod} with $V_{1}=3t$ and $V_{2}=2t$. 
Different colors correspond to different hopping magnitudes.}
\end{figure*}

The next phase that we identify is labelled Kekul\'e phase shown in green color in Fig. \ref{Fig:PD3x3}(a). It has a four-fold degenerate ground state [Fig. \ref{Fig:PD3x3}(c)] at $Q=0 (\times2),4,8$ corresponding to two states at the $\bs{\Gamma}$ point and one state at both $\bs{K}$ and $\bs{K'}$.  As mentioned above, ED of finite systems can only yield precursors of spontaneous symmetry breaking in the thermodynamic limit. This means that if the Kekul\'e order is present, it should appear in all of its linearly independent forms which can be naively counted to be six, depicted in Fig.~\ref{Fig:Kekulepatt}(i).
However, of these six possible Kekul\'e orders, three for each NN hopping distortion $t\pm\delta t$, only four are linearly independent, and the remaining two can be obtained as linear combinations of the other four [see caption in Fig.~\ref{Fig:Kekulepatt}(i)]. 
For example, from the six patterns in Fig 5 (i) we can produce the
second row out of the first row \emph{only} if we use an overall,
homogeneous decrease in hopping. This particular shift one can interpret 
as a linear combination of the second row, i.e. d) + e) + f). 
Therefore the first row plus a linear homogeneous combination
of the second row produces all states in the second row and so only 
four states are independent.
The Kekul\'e bond order is therefore a candidate phase to explain the apparent four fold degenerate ground states in some region of the phase diagram [see Fig.~\ref{Fig:PD3x3}(c)].
The fact that the momenta appear at $\bs{\Gamma}$,  $\bs{K}$ and $\bs{K'}$ supports this scenario since for a unit cell three times larger they all fold into the $\bs{\Gamma}$ point, which means that one can build a hopping perturbation with a periodicity of three unit cells, exactly as the Kekul\'e would need.

In order to further evidence that this is a Kekul\'e phase we explore a construction similar to the charge density wave matrix \eqref{densop}
this time for the hopping amplitudes
\begin{equation}\label{hopop}
t^{mm'}_\bs{r} = \dfrac{1}{\Omega}\sum_{\bs{k}}\left\langle m\right|a^\dagger_{\bs{k}+\left(\bs{Q}_{m}-\bs{Q}_{m'}\right)}b_{\bs{k}}\left|m' \right\rangle e^{i\left(\bs{Q}_{m}-\bs{Q}_{m'}\right)\cdot \bs{r}}
\end{equation} 
with the same notation as above but now for $m=1,2,3,4$ quasi-degenerate ground states. In this case the momentum differences $\bs{Q}_{m}-\bs{Q}_{m'}\in\{0,\bs{K},\bs{K'}\}$ allow for a Kekul\'e bond order. As before, we diagonalize the matrix~\eqref{hopop} and label the system of four eigenvalues and eigenvectors $\lambda_{m}$ and $\bs{v}^{(m)}_{\bs{r}}$. This time, the eigenvectors depend on position. If present, the Kekul\'e bond order will appear as a superposition of the allowed phase factors $e^{i(\bs{Q}_{m}-\bs{Q}_{m'})\cdot\bs{r}}$. We can construct four independent superpositions corresponding to the four eigenvectors such that:
\begin{equation}\label{kekmod}
t^{m}_{\bs{r}}=2\cos(v^{(m)}_{\bs{r},1}+ v^{(m)}_{\bs{r},2}+v^{(m)}_{\bs{r},3}+v^{(m)}_{\bs{r},4}).
\end{equation}  
with $m=1,2,3,4$. When evaluated at the three different links $t^{m}_{\bs{r}},t^{m}_{\bs{r+a}_{1}},t^{m}_{\bs{r+a}_{2}}$ the underlying hopping lattice of this phase is revealed. There are four of such patterns, one for each value of $m$. \\
However, if in the Kekul\'e phase, this procedure will in general produce an arbitrary superposition of all the possible Kekul\'e structures of Fig.~\ref{Fig:Kekulepatt}(i). The four independent superpositions are shown in Fig.~\ref{Fig:Kekulepatt}(ii) where the three different colors represent different bond strengths. Note that each of these patterns has a tripled unit cell periodicity with the right Kekul\'e orders, inherited from the $\bs{r}-$dependent vectors $\bs{v}^{(m)}_{\bs{r}}$. Indeed, by forming linear combinations of these, 
one can obtain all the ``pure" (coherent) Kekul\'e patterns in Fig.~\ref{Fig:Kekulepatt}(i). We find the analysis of this section to be consistent with the presence of the Kekul\'e phase in this part of the phase diagram.
\subsection{Sub-lattice charge modulation (CMs)}
We finally address the last phase that remains to be characterized which we shall name as sublattice charge modulation or CMs 
appearing at the upper left corner of the phase diagram in  Fig.~\ref{Fig:PD3x3}(a).
This phase, unlike the CDW, \emph{does not} correspond to the naive classical strong coupling phase in the corresponding strong coupling limit. Rather, the limit $V_2/t,V_2/V_1\to\infty$ has an extensive classical ground state degeneracy so that quantum corrections will determine the form of the actual ground state for arbitrarily small non-zero values of $t/V_2$ and $V_1/V_2$.  The classical counting yields a $666$ fold degenerate ground state with energy $18V_{2}$ for a $\Omega=3\times 3$ lattice. 
This information serves in fact as a consistency check just as in the CDW case. Indeed, we recover numerically the correct degeneracy and ground state energy in the limit $V_{1}/t\to 0$ and $V_{2}\neq 0$. 

\begin{figure}[h]
\includegraphics[scale=0.2]{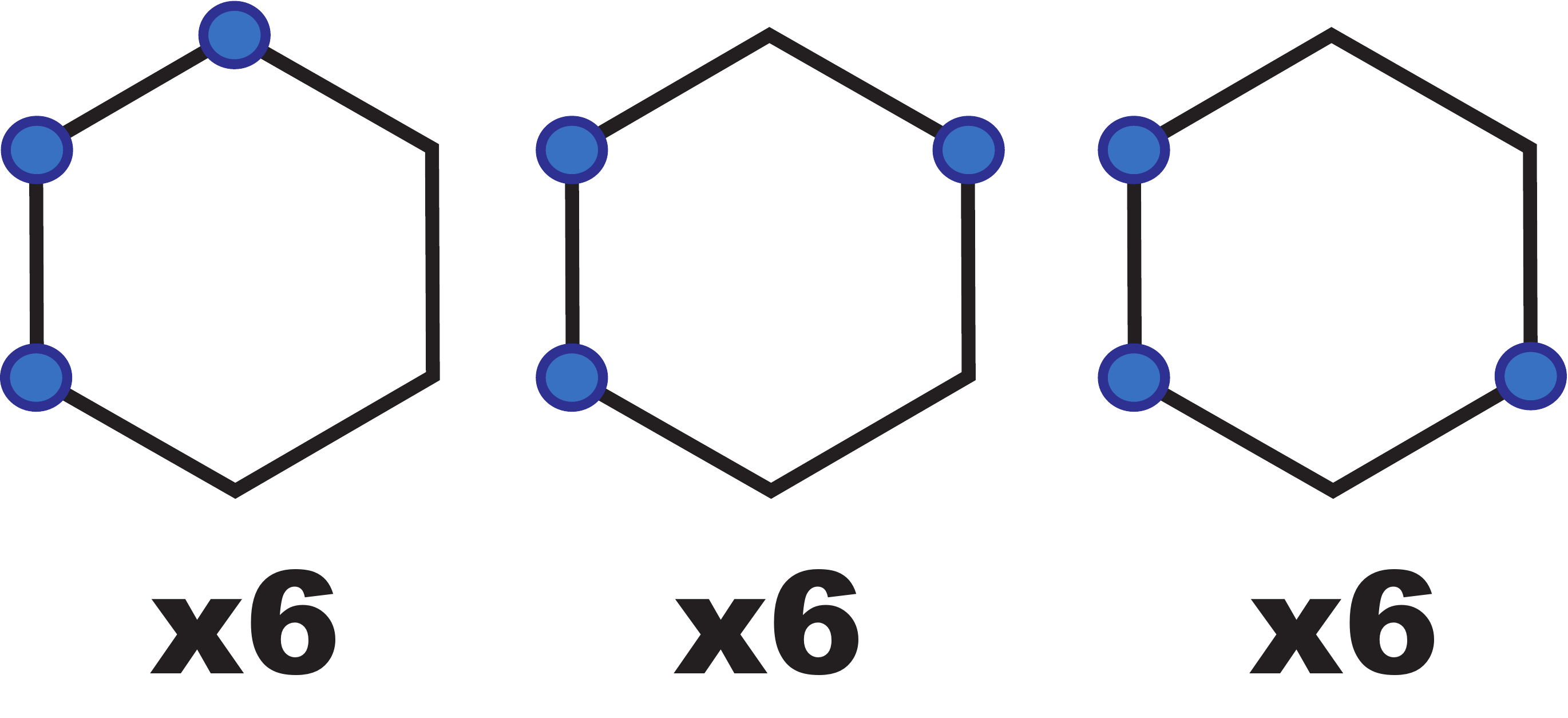}
\caption{\label{Fig:CMspatt} (Color online) CMs patterns with their corresponding degeneracies due 
to a sixfold rotational symmetry.}
\end{figure}

The question then becomes, what phase will be selected by the quantum fluctuations out of the classical ground state manifold.
From a large system with periodic boundary conditions
a natural phase to be expected at half-filling for large $V_{2}/V_{1}$ is that with a charge modulation within the
same sublattice, the CMs phase. This phase, discussed in detail in Ref.~\onlinecite{GCC13}, reduces $V_{2}$ by paying an additional $V_{1}$ cost. Pictorically the state is shown in Fig.~\ref{Fig:CMspatt} where it is evident that it has a degeneracy of 18 because of the rotational symmetry of the Hamiltonian.
We find a phase consistent with this picture at large $V_{2}/V_{1}$ in ED with a quasi-degeneracy of $18$. 
Such a degeneracy slowly becomes more exact as one increases $V_{2}/V_{1}$, although the gap to the excited states also decreases such that in the limit $V_{1}/t\to 0$ and $V_{2}\neq 0$ the strong coupling phase is recovered. Note also that, just as the Kekul\'e , the CMs state has also a tripled unit cell periodicity \cite{GCC13} which is consistent with having the ground states at momenta $\bs{Q}\in\{0,\bs{K},\bs{K'}\}$.
\section{\label{sec:hald}Discussion:}
Despite the small system size, it is remarkable that the phase diagram Fig.~\ref{Fig:PD3x3}(i) resembles very closely the mean-field phase diagram of Ref.~\onlinecite{GCC13} at half-filling shown in Fig.~\ref{Fig:PD3x3} (ii) . We have found, out of the five mean-field phases, all but the CI phase. 

The $\Omega=3\times 3$ lattice studied above is special in that not only it contains the $\bs{K}$ and $\bs{K'}$ which enables clear physical interpretation of the emerging phases, but also it fits phases with a tripled unit cell without frustration, such as the Kekul\'e or the CMs phases. Therefore, such a lattice size has a natural bias towards these phases as compared to the CI phase, which does not break translational symmetries. This might be the reason why the Kekul\'e phase is so prevalent as compared to the mean-field phase diagram in Fig. \ref{Fig:PD3x3} (ii). It is also interesting to note that the Kekule phase shifts to higher values of $V_2$ reducing the region for the charge modulated phase when comparing with the mean field result. Since both phases are favored by the $\Omega=3\times 3$ lattice size this result seems robust.

To investigate further the presence of the CI phase we have studied the $\Omega=3\times 4$ and $4\times 3$ lattices for $V_{1}=0$, where the CI phase is expected to appear from the mean-field analysis at intermediate $V_{2}$. These lattice size frustrates the Kekul\'e and can leave phase space for other phases (such as the CI phase) to appear.

The CI phase for a finite system would appear as a two-fold quasi degenerate ground-state one for each sign of the flux, in a similar way as the CDW shows a two-fold degenerate ground-state corresponding to $\pm m$ or $\mp m$ charge in the A and B sublattices. However, we find no evidence of such signature and thus we conclude that this phase is absent also from the ED of $\Omega=3\times 4$ and $4\times 3$ lattices. Taking the $3\times 4$ as an example, the spectra along the $V_{2}$ line shows first a single ground-state at $Q=6$ for low $V_{2}$, as expected for the trivial SM phase just by adding all the non-interacting momenta in Fig. \ref{Fig:nonintbands}(b). At $V_{2}\sim 7t $ the gap closes and reopens with a six fold quasi-degenerate ground-state. The lowest pair of these states 
lie at $Q=6$, as would be expected for the CI phase. Therefore, it remains an open question whether this six-fold degeneracy
becomes two-fold by increasing the lattice size, which could in 
principle lead in the thermodynamic limit to the appearance of 
the Haldane phase.

Finally we comment on a different route towards achieving interaction driven topological phases in the extended Hubbard model on the honeycomb lattice which involves moving away from half-filling. An example of these topological phases were shown to appear at higher fillings from a mean-field calculation in Ref \onlinecite{CGV11,GCC13}. These are generalizations 
of the Haldane phase at fillings $\nu\gtrsim 2/3$ and with a tripled unit cell, which could also
be present via ED. However, identifying these phases by characterizing the ground state properties
from ED is challenging due to band folding.\\
\section{\label{sec:conc}Conclusions}
We have investigated the effect of extended Hubbard interactions on spinless electrons on the honeycomb lattice at half filling
via exact diagonalization (ED). We have found that four out of the five predicted mean-field phases are present. 
These are the semi-metal (SM), CDW,
the Kekul\'e bond order and the sublattice charge modulation (CMs) phases. First, we have shown that the six-fold degeneracy of the
SM ground-state can be understood entirely from the non-interacting band structure. For the CDW phase we have proven that it is connected to the strong coupling phase at $V_{1}/t\to \infty$,$V_{2}=0$ and we have characterized it finding a finite sublattice charge imbalance through the charge order correlation function, the hallmark for the CDW phase. The two-fold degeneracy is a sign of the
two possible orders that the system can choose in the thermodynamic limit by spontaneously breaking the sublattice symmetry.
Similarly, we have disentangled the Kekul\'e bond order phase by calculating the bond order correlation function which reveals the underlying superposition of four independent Kekul\'e patterns which conform the fourfold quasi-degenerate ground-state.
Finally, we have argued that for $V_{2}>V_{1}$ the CMs phase is expected to have a 18-fold degeneracy favoured by the costly NNN interaction, which is consistent with what we observed in ED.

Importantly, the fact that the discussed phases appear both in ED and in mean-field suggests that they are stable up to the thermodynamic limit. The appearance of the Kekul\'e phase dominating a wide region of the phase diagram opens up the possibility of realizing this exotic phase in cold atoms with a scheme along the lines of Ref.~\onlinecite{DMM12}. Despite the fact that we have not found evidence for the Chern Insulator (CI) phase, it is still possible that it is realizable in the thermodynamic limit. Different approaches such as cluster mean-field \cite{DH13} can also prove useful to ascertain the presence of the CI phase. We hope that the conclusions of the present work will motivate further explorations of the extended Hubbard model on the honeycomb lattice. \\

\emph{Note added:} 
During the completion of this work we learned from a complementary analysis \cite{DH13} that focuses on the line  $V_{2}\neq 0$, $V_{1}=0$ in the phase diagram. The results are consistent with those presented here, in particular, with the absence of the CI phase and the appearance of the CMs state.

\begin{acknowledgments}
We acknowledge conversations with M. A. H. Vozmediano, F. de Juan,
A. Cortijo, C. Mudry. 
We also thank M. Hohenadler and M. Daghofer for sharing their results
prior to publication. 
This research was supported in part by the Spanish MECD grants 
FIS2011-23713, FIS2011-29689, PIB2010BZ-00512 and by the Swiss National Science Foundation.
\end{acknowledgments}



\end{document}